\documentclass{article}

\input{tcilatex}
\begin{document}

\title{Tensor Lagrangians, Lagrangians equivalent to the Hamilton-Jacobi
equation and relativistic dynamics}
\author{Alexander Gersten \\
Department of Physics\\
Ben-Gurion University of the Negev\\
84105 Beer-Sheva, Israel\\
alex.gersten@gmail.com}
\maketitle

\begin{abstract}
We deal with Lagrangians which are not the standard scalar ones. We present
a short review of tensor Lagrangians, which generate massless free fields
and the Dirac field, as well as vector and pseudovector Lagrangians for the
electric and magnetic fields of Maxwell's equations with sources. We
introduce and analyse Lagrangians which are equivalent to the
Hamilton-Jacobi equation and recast them to relativistic equations.

\textbf{Key Words, }scalar Lagrangians, tensor Lagrangians, Hamilton-Jacobi
equation, relativistic dynamics
\end{abstract}

\bigskip (In Foundations of Physics: Found Phys (2011) 41: 88--98)

\section{\protect\bigskip \protect\bigskip Introduction}

Maxwell's equations are a peculiar case for which there does not exist a
scalar Lagrangian generating the equations of the electric and magnetic
fields. There exist scalar Lagrangians generating equations for the
electromagnetic potentials. Maxwell's equations played a very important role
in the development of theoretical physics. They played a crucial role in the
development of special relativity as they were Lorentz invariant. They even
encompass the one photon quantum equation$^{\cite{ger2},\cite{ger3},\cite%
{bial},\cite{esp}}$. One would expect to see the Planck constant in a
quantum equation, but we have observed that the Planck constant cancels out
in massless free field (homogeneous) equations$^{\cite{ger2},\cite{ger4}}$;
that explains its absence in Maxwell's equations.

Maxwell's equations have large amount of important symmetries. Most of them
were described in the book of Fushchych and Nikitin$^{\cite{fush1}}$. We
have also developed methods for finding symmetries of Maxwell's equations$^{%
\cite{ger1}}$, but now we realize that they were a drop in the infinite sea
of symmetries.

The Lagrangian formulation of the interaction of charged particles with the
electromagnetic potentials was crucial in the development of classical and
quantum electrodynamics. The Lagrangian formalism is extremely important in
theoretical physics, yet it is not uniquely defined and not always
applicable. Our aim is to bring forward new approaches and new perspectives.
In the present paper we deal with Lagrangians which are not the standard
scalar Lagrangians.

Tensor Lagrangians were derived in the past for all massless free fields and
the Dirac field$^{\cite{morgan}}$.Vector Lagrangians were constructed for
the electric and magnetic fields of the Maxwell equations with currents$^{%
\cite{Sudbery},\cite{fush}}$. These Lagrangians gave rise to a large number
of conserved currents. As these Lagrangians are not well known, a short
review is given in Sec. 2.

In Sec. 3 Lagrangians equivalent to the Hamilton-Jacobi equation (HJE) will
be presented. In addition to our previous work$^{\cite{ger}}$, the
relativistic dynamics will be exposed in more details. Similarly to the
HJE's, these Lagrangians generate all possible trajectories. The merit of
this approach is that it generates equations for the first integrals of the
HJE namely the conjugate momenta and energy, thus it can be thought as a
shortcut to the HJE. The main difference between standard classical mechanics%
$^{\cite{gold}}$ and our approach is that in the standard classical
mechanics the coordinates and momenta are functions of time only. In our
approach the momenta are functions of the coordinates and the coordinates
are functions of time. Thus the coordinate-dependent momenta can be
considered as the fields of classical mechanics.

Summary and conclusions will be given in Sec. 4.

\section{Tensor Lagrangians}

In this section a short review of tensor Lagrangians for free fields and
vector Lagrangians for Maxwell equations with sources will be given.
Repeated indices will induce the summation convention.

\subsection{Tensor Lagrangians for free massless fields and for the Dirac
field}

Morgan and Joseph$^{\cite{morgan}}$ considered the Pauli-Fiertz equation$^{%
\cite{pau},\cite{cor}}$ for a massless free field of spin $s$, represented
by a completely symmetric spinor $\varphi _{A_{1}...A_{2s}}$ with $2s$
indices

\begin{equation}
\partial ^{A_{1}\dot{B}}\varphi _{A_{1}...A_{2s}}=0,\quad or\quad \partial
^{A\dot{B}_{1}}\varphi _{\dot{B}_{1}...\dot{B}_{2s}}=0  \label{1}
\end{equation}%
where $\partial ^{A\dot{B}}$ is the derivative operator in spinor form. They
found that the Pauli-Fierz equations can be obtained from the tensor
Lagrangian

\begin{equation}
L_{A_{2}...A_{2s}\dot{B}_{2}...\dot{B}_{2s}}=\frac{i}{2}\left[ \varphi
_{A_{1}...A_{2s}}\partial ^{A_{1}\dot{B}_{1}}\varphi _{\dot{B}_{1}...\dot{B}%
_{2s}}-\left( \partial ^{A_{1}\dot{B}_{1}}\varphi _{A_{1}...A_{2s}}\right)
\varphi _{\dot{B}_{1}...\dot{B}_{2s}}\right] .  \label{2}
\end{equation}

The Dirac equation is obtained from the Lagrangian

\begin{equation}
L=\frac{i}{2}\left[ \varphi _{A}\partial ^{A\dot{B}}\varphi _{\dot{B}%
}-\left( \partial ^{A\dot{B}}\varphi _{A}\right) \varphi _{\dot{B}}\right] +%
\left[ \frac{i}{2}m(\chi ^{A}\partial _{\mu }\varphi _{A}+\varphi _{\dot{B}%
}\partial _{\mu }\chi ^{\dot{B}})+h.c.\right] .  \label{2a}
\end{equation}

\subsection{Vector and pseudovector Lagrangians for the electromagnetic
field with sources}

For the full set of Maxwell's equations

\begin{equation}
\partial _{\mu }F^{\mu \nu }=j^{\nu },\quad \partial _{\mu }\bar{F}^{\mu \nu
}=0,  \label{3}
\end{equation}%
where $F_{\mu \nu }$ is the electromagnetic field tensor and $\bar{F}^{\mu
\nu }$ its dual, Sudbery$^{\cite{Sudbery}}$ found the following 4-vector
Lagrangian%
\begin{equation}
\mathit{L}_{\alpha }=\bar{F}^{\mu \nu }\partial _{\nu }F_{\mu \alpha
}-F^{\mu \nu }\partial _{\nu }\bar{F}_{\mu \alpha }-2\bar{F}_{\alpha \mu
}j^{\mu }  \label{4}
\end{equation}

Fushchych, Krivskiy and Simulik$^{\cite{fush}}$ found an infinite number of
Lagrangians which generate Eqs.(\ref{3}). Using their notation

\begin{equation}
Q_{\mu }\equiv \partial _{\nu }F,\quad R_{\mu }\equiv \partial _{\nu }\bar{F}%
_{\mu \nu },  \label{5}
\end{equation}%
Eqs.(\ref{3}) become

\begin{equation}
Q_{\mu }=j_{\mu },\quad R_{\mu }=0,\quad \mu =0,1,2,3,  \label{6}
\end{equation}%
They consider the tensor $T_{\mu \rho \sigma }$ and pseudotensor $T_{\mu
\rho \sigma }^{\prime }$ with respect to the total Poincar\'{e} group $%
\tilde{P}(1,3)$

\begin{equation}
T_{\mu \rho \sigma }\equiv a[g_{\mu \rho }(Q_{\sigma }-j_{\sigma })-g_{\mu
\sigma }(Q_{\rho }-j_{\rho })]+b\epsilon _{\mu \nu \rho \sigma }R^{\nu },
\label{7}
\end{equation}

\begin{equation}
T_{\mu \rho \sigma }^{\prime }\equiv a^{\prime }(g_{\mu \rho }R_{\sigma
}-g_{\mu \sigma }R_{\rho })+b^{\prime }\epsilon _{\mu \nu \rho \sigma
}(Q_{\nu }-j_{\nu }),  \label{8}
\end{equation}%
where $a,b,a^{\prime },b^{\prime }$ are constant coefficients and $\epsilon
^{\mu \nu \rho \sigma }$ is a completely antisymmetric unit tensor, $%
\epsilon ^{0123}=1$. They proved that for any $ab=0=a^{\prime }b^{\prime }$
each of the set of equations

\begin{equation}
T_{\mu \rho \sigma }=0,  \label{9}
\end{equation}

\begin{equation}
T_{\mu \rho \sigma }^{\prime }=0,  \label{10}
\end{equation}%
is equivalent to the Maxwell equations Eqs.(\ref{3}). They considered the
following Lagrangian:

\begin{eqnarray}
L_{\mu } &=&a_{1}F_{\mu \nu }Q^{\nu }+a_{2}F_{\mu \nu }\bar{R}^{\nu
}+a_{3}\epsilon F_{\mu \nu }R^{\nu }+a_{4}\epsilon F_{\mu \nu }\bar{Q}^{\nu
}+a_{5}\bar{F}_{\mu \nu }Q^{\nu }  \nonumber \\
&&+a_{6}\bar{F}_{\mu \nu }R^{\nu }+a_{7}\epsilon \bar{F}_{\mu \nu }\bar{R}%
^{\nu }+a_{8}\epsilon \bar{F}_{\mu \nu }Q^{\nu }+(q_{1}F_{\mu \nu
}+q_{2}\epsilon \bar{F}_{\mu \nu })j^{\nu },  \label{11}
\end{eqnarray}%
where 
\[
\bar{Q}_{\mu }\equiv \partial _{\nu }\bar{F}_{\mu \nu },\quad \bar{R}\mu
\equiv \partial _{\nu }\epsilon \bar{F}_{\mu \nu },\quad \epsilon \bar{F}%
_{\mu \nu }\equiv \frac{1}{2}\epsilon ^{\mu \nu \rho \sigma }\bar{F}_{\rho
\sigma } 
\]%
They proved that Maxwell equations are generated from the Lagrangian Eqs.(%
\ref{11}) if and only if the following conditions on the coefficients are

\begin{eqnarray}
a_{8}-a_{2} &=&a=-b^{\prime }=-q_{1}\equiv -q=0,  \nonumber \\
a_{6}-a_{4} &=&a^{\prime }=-b\neq 0,  \nonumber \\
a_{1}-a_{3}-a_{6}-a_{8} &=&a_{2}+a_{4}+a_{5}-a_{7}=0.  \label{12}
\end{eqnarray}

\section{\protect\bigskip Lagrangians equivalent to the Hamilton-Jacobi
equation}

We will start with nonrelativistic dynamics and consider n-dimensional
Lagrangians

\begin{equation}
L(x_{1},...,x_{n},\dot{x}_{1},...,\dot{x}_{n},t)\equiv L(x,\dot{x},t),\quad 
\dot{x}=\frac{dx}{dt}.  \label{12a}
\end{equation}%
Hamilton's principle of least action

\begin{equation}
S(x)=\int_{t_{1}}^{t_{2}}L(x,\dot{x}_{,}t)dt=\min ,  \label{13}
\end{equation}%
where $S$ is the action, leads to the Euler-Lagrange equations%
\begin{equation}
\frac{\partial L}{\partial x_{i}}=\frac{d}{dt}\frac{\partial L}{\partial 
\dot{x}_{i}},\quad i=1,...,n  \label{14}
\end{equation}%
or to the Hamilton's equations%
\begin{equation}
\frac{\partial H(x_{,}p,t)}{\partial x_{k}}=-\dot{p}_{k},\quad \frac{%
\partial H(x,p,t)}{\partial p_{k}}=\dot{x}_{k},\quad k=1,...,n,  \label{15}
\end{equation}%
where $H$ is the Hamiltonian and $p_{i}$ the momenta. The solutions are
trajectories $x_{i}(t)$.

Let us adopt another approach. Let us change Eq.(\ref{13}) to%
\begin{equation}
S(x,t)=\int_{t_{0}}^{t}L(x,\dot{x}_{,}t^{\prime })dt^{\prime }=\min ,
\label{16}
\end{equation}%
and go to the Hamiltonian formalism replacing $L$ with a Lagrangian $%
L_{_{HJ}}$ which we will show later on, will generate the HJE.%
\begin{equation}
L_{_{HJ}}(x,\dot{x},t)=p_{i}(x,t)\dot{x}_{i}-H(x,p(x,t),t),\quad i=1,...,n,
\label{17}
\end{equation}%
From Eq.(\ref{16}) and Eq.(\ref{17}) we obtain%
\begin{equation}
dS(x,t)=L_{_{HJ}}dt=p_{i}(x,t)dx_{i}-H(x,p(x,t),t)dt,\quad i=1,...,n,
\label{18}
\end{equation}%
from which the HJE immediately results%
\begin{equation}
\frac{\partial S(x,t)}{\partial x_{j}}=p_{j}(x,t),\quad \frac{\partial S(x,t)%
}{\partial t}=-H(x,p,t).  \label{19}
\end{equation}

The Euler-Lagrange equations for the Lagrangian $L_{_{HJ}}$ of Eq.(\ref{17})
are%
\begin{equation}
\frac{\partial L_{_{HJ}}}{\partial x_{j}}-\frac{d}{dt}\frac{\partial
L_{_{HJ}}}{\partial \dot{x}_{j}}=\left( \frac{\partial p_{i}}{\partial x_{j}}%
\dot{x}_{i}-\frac{\partial H}{\partial p_{i}}\frac{\partial p_{i}}{\partial
x_{j}}\right) -\frac{\partial H}{\partial x_{j}}-\frac{d}{dt}p_{j}.
\label{191}
\end{equation}%
Later on we will also use the relations%
\begin{equation}
\frac{\partial L_{_{HJ}}}{\partial x_{j}}=\frac{d}{dt}\frac{\partial
L_{_{HJ}}}{\partial \dot{x}_{j}}=\frac{d}{dt}p_{j}\equiv \dot{p}_{j},\quad 
\frac{\partial L_{_{HJ}}}{\partial \dot{x}_{j}}=p_{j},\quad \frac{d}{dt}%
p_{j}=\frac{\partial p_{j}}{\partial x_{k}}\dot{x}_{k}+\frac{\partial p_{j}}{%
\partial t},  \label{192}
\end{equation}%
and%
\begin{equation}
dL_{_{HJ}}(x,\dot{x},t)=\frac{\partial L_{_{HJ}}}{\partial x_{i}}dx_{i}+%
\frac{\partial L_{_{HJ}}}{\partial \dot{x}_{i}}d\dot{x}_{i}+\frac{\partial
L_{_{HJ}}}{\partial t}dt=\dot{p}_{i}dx_{i}+p_{i}d\dot{x}_{i}+\frac{\partial
L_{_{HJ}}}{\partial t}dt,  \label{193}
\end{equation}%
in order to find a non-linear equation for the momenta.

\subsection{The general case}

Using Eq.\ref{17}) and Eq.(\ref{193}) we have%
\begin{equation}
dH(x,p(x,t),t)=\dot{x}_{i}dp_{i}+p_{i}d\dot{x}_{i}-dL_{_{HJ}}=\dot{x}%
_{i}dp_{i}-\frac{\partial L_{_{HJ}}}{\partial x_{i}}dx_{i}-\frac{\partial
L_{_{HJ}}}{\partial t}dt,  \label{221}
\end{equation}%
from which we deduce that%
\begin{equation}
\frac{\partial H}{\partial x_{i}}=-\frac{\partial L_{_{HJ}}}{\partial x_{i}}%
,\quad \frac{\partial H}{\partial p_{i}}=\dot{x}_{i},\quad \frac{\partial H}{%
\partial t}=-\frac{\partial L_{_{HJ}}}{\partial t},  \label{222}
\end{equation}%
but

\[
\frac{\partial H}{\partial x_{i}}=-\frac{\partial L_{_{HJ}}}{\partial x_{i}}%
=-\frac{d}{dt}p_{i}=-\dot{p}_{i}, 
\]%
and Hamilton-like equations are obtained%
\begin{equation}
\frac{\partial H(x,p(x,t),t)}{\partial x_{i}}=-\dot{p}_{i}(x,t),\quad \frac{%
\partial H(x,p(x,t),t)}{\partial p_{i}}=\dot{x}_{i}.  \label{224}
\end{equation}%
Noting that%
\begin{equation}
\dot{p}_{i}(x,t)=\frac{d}{dt}p_{i}(x,t)=\frac{\partial p_{i}(x,t)}{\partial
x_{j}}\dot{x}_{j}+\frac{\partial p_{i}(x,t)}{\partial t},  \label{225}
\end{equation}%
and substituting the expression for $\dot{x}_{i}$ in Eq.(\ref{224}) we
obtain a non-linear equation for the momenta%
\begin{equation}
\frac{\partial H(x,p(x,t),t)}{\partial x_{i}}+\frac{\partial p_{i}(x,t)}{%
\partial x_{j}}\frac{\partial H(x,p(x,t),t)}{\partial p_{j}}+\frac{\partial
p_{i}(x,t)}{\partial t}=0.  \label{226}
\end{equation}

Let us look for first integrals of Eq.(\ref{226}). Let us consider%
\begin{equation}
dH(x,p(x,t),t)=\dot{x}_{i}dp_{i}+\frac{\partial H}{\partial x_{i}}dx_{i}+%
\frac{\partial H}{\partial t}dt=\dot{x}_{i}dp_{i}-\dot{p}_{i}dx_{i}+\frac{%
\partial H}{\partial t}dt,  \label{227}
\end{equation}%
and divide Eq.(\ref{227}) by $dt$, we obtain%
\[
\frac{dH}{dt}=\dot{x}_{i}\dot{p}_{i}-\dot{p}_{i}\dot{x}_{i}+\frac{\partial H%
}{\partial t}=\frac{\partial H}{\partial t} 
\]%
i.e.%
\begin{equation}
\frac{dH(x,p(x,t),t)}{dt}=\frac{\partial H(x,p(x,t),t)}{\partial t}.
\label{228}
\end{equation}

\subsection{The case of time independent Hamiltonian}

If the Hamiltonian is time independent, the solutions of the HJE have the
form%
\begin{equation}
S(x,t)=W(x)-Et+C  \label{228a}
\end{equation}%
where $E$ and $C$ are constants. The HJE (Eq.(\ref{19}))\ \ becomes 
\begin{equation}
\frac{\partial S(x,t)}{\partial x_{j}}=\frac{\partial W(x)}{\partial x_{j}}%
=p_{j}(x),\quad H(x,\nabla S)=E,  \label{228b}
\end{equation}%
i.e. the momenta are functions of the coordinates only, describing all
possible trajectories with the same total energy $E$. The Hamiltonian, using
Eq.(\ref{228}) becomes a constant of motion%
\begin{equation}
\frac{dH(x,p(x))}{dt}=0.  \label{229}
\end{equation}%
From Eq.(228b) we have

\begin{equation}
\frac{\partial p_{i}}{\partial x_{k}}-\frac{\partial p_{k}}{\partial x_{i}}=%
\frac{\partial ^{2}S}{\partial x_{k}\partial x_{i}}-\frac{\partial ^{2}S}{%
\partial x_{i}\partial x_{k}}=0  \label{229a}
\end{equation}

Let us show that%
\begin{equation}
\frac{dH(x,p(x))}{dx_{k}}=0.  \label{230}
\end{equation}%
Indeed, using Eq.(\ref{227}), Eq.(\ref{229a}) and Eq.(\ref{228b}) we have%
\[
\frac{dH}{dx_{k}}=\dot{x}_{i}\frac{dp_{i}}{dx_{k}}-\frac{dp_{k}}{dx_{i}}\dot{%
x}_{i}=\dot{x}_{i}\left( \frac{dp_{i}}{dx_{k}}-\frac{dp_{k}}{dx_{i}}\right) =%
\dot{x}_{i}\left( \frac{\partial p_{i}}{\partial x_{k}}-\frac{\partial p_{k}%
}{\partial x_{i}}\right) =0 
\]%
Hence we can deduce that the Hamiltonian is the constant of motion for all
trajectories%
\begin{equation}
H(x,p(x))=E.  \label{23}
\end{equation}

\bigskip The non linear equation for the momenta Eq.(\ref{226}) becomes now

\begin{equation}
\frac{\partial H(x,p(x))}{\partial x_{i}}+\frac{\partial p_{i}(x)}{\partial
x_{j}}\frac{\partial H(x,p(x))}{\partial p_{j}}=0  \label{23a}
\end{equation}

From equations Eq.(\ref{23a}), Eq.(\ref{229a}) and Eq.(\ref{23}) one can
find all the trajectories $p_{i}(x)$ having the same energy $E$.

\subsubsection{Example 1}

Let us consider, as a special case the Hamiltonian in one dimension

\begin{equation}
H\left( x,p\left( x\right) \right) =\frac{p^{2}(x)}{2m}+V(x)  \label{28}
\end{equation}%
We can start from Eq.(\ref{23}), but for better illustration we use Eq. (\ref%
{226}), which becomes

\begin{equation}
\frac{p}{m}\frac{dp}{dx}=-\frac{dV}{dx},  \label{9a}
\end{equation}%
Eq.(\ref{9a}) can be recast into 
\begin{equation}
\frac{d}{dx}\left( \frac{p^{2}}{2m}+V\right) =0,  \label{10x}
\end{equation}%
hence $H$ is a constant in $x$ and may depend only on time

\begin{equation}
H\left( x,p\left( x\right) \right) =\frac{p^{2}\left( x\right) }{2m}%
+V(x)=E(t).  \label{10a}
\end{equation}%
Moreover if $H$ does not depend explicitly on time

\begin{equation}
\frac{\partial H}{\partial t}=0\Longrightarrow E=const.,  \label{10b}
\end{equation}%
and the total energy is conserved (of course we could have started from this
point using Eq.(\ref{23}))

\begin{equation}
\frac{p^{2}\left( x\right) }{2m}+V\left( x\right) =E.  \label{10c}
\end{equation}%
The momentum as a (double valued) field is given by

\begin{equation}
p(x)=\pm \sqrt{2mE-V(x)}.  \label{10d}
\end{equation}%
Usually the sign ambiguity can be resolved using physical considerations.
The trajectories may be obtained from Eq.(\ref{224}), which in our case is%
\begin{equation}
\dot{x}=\frac{p\left( x\right) }{m},  \label{10f}
\end{equation}%
and by integrating

\begin{equation}
dt=\frac{m}{p\left( x\right) }dx=\frac{\pm mdx}{\sqrt{2mE-V(x)}}.
\label{10g}
\end{equation}%
The action can be evaluated from Eq.(\ref{19}) and Eq.(\ref{228a})

\begin{equation}
S(x,t,;x_{0},t_{0})=\pm \int_{x_{0}}^{x}\sqrt{2mE-V(x^{\prime })}dx^{\prime
}-E(t-t_{0}).  \label{10e}
\end{equation}

\subsection{Relativistic dynamics of one particle}

\bigskip We will show that Eq.(\ref{18}), although derived from
nonrelativistic mechanics, can be considered as a relativistic one if the
momenta and energy (the Hamiltonian) belong to the same 4-vector.

If we consider Eq.(\ref{18}) as a relation between two Lorentz scalars, then
the relativistic Lagrangian can be defined as (using the metric (+,-,-,-))%
\begin{eqnarray}
L_{rel} &=&\frac{dS}{d\tau }=-p_{\mu }\frac{dx^{\mu }}{d\tau }=-p_{i}(x)%
\frac{dx_{i}}{d\tau }-H(x,p)\frac{dt}{d\tau };\quad  \label{31} \\
\mu &=&0,1,2,3;\quad i=1,2,3;\quad x_{0}=ct;\quad p_{0}=H/c;\quad \mathring{x%
}^{i}=\frac{dx^{i}}{d\tau },  \nonumber
\end{eqnarray}

\bigskip where $\tau $ is the proper time.

The action to be minimized is%
\begin{equation}
S\left( x,\tau \right) =\int_{\tau _{0}}^{\tau }L_{rel}\left( x,\frac{dx}{%
d\tau },\tau ^{\prime }\right) d\tau ^{\prime }=\int^{t}L_{rel}\frac{d\tau
^{\prime }}{dt^{\prime }}dt^{\prime }=\int^{t}\frac{dS}{dt^{\prime }}%
dt^{\prime }=\int_{t_{0}}^{t}L_{_{HJ}}dt^{\prime }=S\left( x,t\right) ,
\label{31a}
\end{equation}%
i.e. the same as in the nonrelativistic case, provided $p_{i}\left( x\right) 
$ and $H(x,p(x))/c$ are components of the same 4-vector. Therefore all
equations previously derived in this section can be used for relativistic
dynamics with the Lagrangian

\begin{equation}
L_{_{HJ}}=-p_{i}(x)\dot{x}^{i}-H(x,p).  \label{31b}
\end{equation}

The equations are not covariant but relativistically invariant.

\subsubsection{Relativistic charged particle in a static electromagnetic
field}

Barut$^{[\cite{Barut}]}$ has shown that a proper Hamiltonian for this case
(which is the energy) is

\begin{equation}
H(x,p)=e\varphi (x)+c\left[ [p_{i}(x)-\frac{e}{c}A_{i}(x)][p_{i}(x)-\frac{e}{%
c}A_{i}(x)]+m^{2}c^{2}\right] ^{%
{\frac12}%
},  \label{31c}
\end{equation}%
where $\varphi (x)$ and $A_{i}(x)$ are the scalar and vector electromagnetic
potentials respectively. In order to find all the trajectories of the
momenta with a constant energy $E$ one has to solve Eq.(\ref{23a}) and Eq.(%
\ref{23}).

\subsubsection{Example 2, free particle with mass m}

The (time independent) Hamiltonian $H$ is

\begin{equation}
H/c=\sqrt{m^{2}c^{2}+p_{i}\left( x\right) p_{i}\left( x\right) },  \label{40}
\end{equation}%
Using Eq.(\ref{23}) we find that the trajectories with constant energy $E$
satisfy

\begin{equation}
p_{i}\left( x\right) p_{i}\left( x\right) =\frac{E^{2}}{c^{2}}-m^{2}c^{2}.
\label{41}
\end{equation}%
From Eq.(\ref{230}) we have

\begin{equation}
\frac{dH}{dx_{j}}=\frac{cp_{i}\left( x\right) \frac{dp_{i}\left( x\right) }{%
dx_{j}}}{\sqrt{m^{2}c^{2}+p_{i}\left( x\right) p_{i}\left( x\right) }}%
=0;\quad p_{i}\left( x\right) \frac{dp_{i}\left( x\right) }{dx_{j}}=0,
\label{42}
\end{equation}%
and from Eq.(\ref{42}) one can deduce that

\begin{equation}
\frac{dp_{i}\left( x\right) }{dx_{j}}=0,  \label{43}
\end{equation}%
i.e. all the trajectories are straight lines with the constraint given by
Eq.(\ref{41}). In order to find a particular trajectory one has to give
initial conditions and solve the Hamilton equation (Eq.(\ref{224}))

\[
\frac{\partial H(x,p(x))}{\partial p_{i}}=\dot{x}_{i}\ , 
\]%
and use the constraint as given by Eq.(\ref{41}).

\subsubsection{Example 3, relativistic charged particle in a static electric
field}

\begin{equation}
\frac{\partial H(x,p(x))}{\partial x_{i}}+\frac{\partial p_{i}(x)}{\partial
x_{j}}\frac{\partial H(x,p(x))}{\partial p_{j}}=e\frac{\partial \varphi (x)}{%
\partial x_{i}}+c\frac{\partial p_{i}(x)}{\partial x_{j}}\frac{p^{j}(x)}{%
\left[ (p_{i}(x)p^{i}(x)+m^{2}c^{2}\right] ^{%
{\frac12}%
}}=0  \label{31h}
\end{equation}%
and

\begin{equation}
e\varphi (x)+c\left[ p_{i}(x)(p^{i}(x)+m_{0}^{2}c^{2}\right] ^{%
{\frac12}%
}=E,  \label{31e}
\end{equation}%
combining the two equations we find

\begin{equation}
c^{2}\frac{\partial p_{i}(x)}{\partial x_{j}}p^{j}(x)=e\frac{\partial
\varphi (x)}{\partial x_{i}}[e\varphi (x)-E],\quad i=1,2,3  \label{31f}
\end{equation}%
The absolute value of the momentum can be found from Eq.(\ref{31e}), namely

\begin{equation}
c^{2}p_{i}(x)p^{i}(x)=\left( E-e\varphi (x)\right) ^{2}-m_{0}^{2}c^{4}.
\label{31j}
\end{equation}%
The components of the momentum can be found by solving Eq.(\ref{31f}) with
initial conditions. An exact solution for an arbitrary potential can be
given if it depends on one coordinate only, for instance

\[
\varphi (x)=\varphi (x_{1}). 
\]%
In this case from Eq.(\ref{31f}) we find

\[
\frac{\partial p_{2}(x)}{\partial x_{j}}=\frac{\partial p_{3}(x)}{\partial
x_{j}}=0;\quad j=1,2,3;\quad p_{2}(x)=P_{2}=const;\quad
p_{3}(x)=P_{3}=const.\quad 
\]%
and from Eq.(\ref{31j})

\[
cp_{1}(x)=\pm \lbrack \left( E-e\varphi (x_{1})\right)
^{2}-m_{0}^{2}c^{4}-c^{2}P_{2}^{2}-c^{2}P_{3}^{2}]^{%
{\frac12}%
}. 
\]

\section{Summary and conclusions}

In the present paper we have dealt with Lagrangians which are not the
standard scalar Lagrangians. It is well known that the Maxwell equations for
the electric and magnetic fields can not be derived from a scalar
Lagrangian. Only the equations for the vector potentials can be derived from
a scalar one.

We gave a short review of tensor Lagrangians applied for free fields and the
Dirac field$^{\cite{morgan}},$ and vector and pseudovector Lagrangians
generating the Maxwell' equations$^{\cite{Sudbery},\cite{fush}}$. In the
cited papers the symmetries of the Lagrangians were used to derive new and
even an infinite amount of conserved currents.

In Sec. 3 Lagrangians equivalent to the Hamilton-Jacobi equation (HJE) were
presented in addition to our previous work$^{\cite{ger}}$, emphasizing the
relativistic dynamics. Similarly to the HJE's, these Lagrangians generate
all possible trajectories. We have shown that a Lagrangian of the form as
given in Eq.(\ref{17}), generates the HJE. We have found the equations of
the momentum field, they are given in Eq.(\ref{226}).The advantage of this
approach is that it generates equations for the first integrals of the HJE
namely the conjugate momenta and energy, without solving the HJE. The case
of time independent Hamiltonian is dealt in subsection 3.2. We have shown
that the solutions (Eq.(\ref{229}) and Eq.(\ref{230})) satisfy $\frac{%
dH(x,p(x))}{dt}=\frac{dH(x,p(x))}{dx_{k}}=0,$ and the energy $E$ is the
constant of motion for all trajectories of the family $H(x,p(x))=E,$ which
is the first integral of the momentum field equations Eq.(\ref{226}) for
time independent Hamiltonians.

The main difference between standard classical mechanics$^{\cite{gold}}$ and
our approach is that in the standard classical mechanics the coordinates and
momenta are functions of time only. In our approach the momenta are
functions of the coordinates for all possible trajectories.

In subsection 3.3 \ we have dealt with relativistic dynamics and we have
shown that the non-relativistic formalism can be used provided the momenta
and (the equall to energy) Hamiltonian belong to the same 4-vector.

\ \

\end{document}